\documentclass[twocolumn,aps,prl,floatfix,superscriptaddress]{revtex4-1}
\usepackage{hyperref}
\usepackage[utf8]{inputenc}

\usepackage{graphicx}
\usepackage{dcolumn}
\usepackage{amsmath}
\usepackage{supertabular}
\usepackage{multirow}

\usepackage{color}
\usepackage{cancel}
\usepackage{ulem}
\usepackage{array}
\usepackage{tabularx}

\usepackage{amsmath}
\usepackage{amssymb}
\usepackage{xcolor}

\begin{document}

\preprint{APS/123-QED}

\title{
Measurement of Kaon Directed Flow \\ 
in Au+Au Collisions in the High Baryon Density Region
%at $\sqrt{s_{\text{NN}}}$ = 3.0, 3.2, 3.5 and 3.9 GeV
}

\bigskip

\affiliation{Academia Sinica, Nankang, 115, Taipei}
\affiliation{Abilene Christian University, Abilene, Texas   79699}
\affiliation{Alikhanov Institute for Theoretical and Experimental Physics NRC "Kurchatov Institute", Moscow 117218}
\affiliation{Argonne National Laboratory, Argonne, Illinois 60439}
\affiliation{American University in Cairo, New Cairo 11835, Egypt}
\affiliation{Ball State University, Muncie, Indiana, 47306}
\affiliation{Brookhaven National Laboratory, Upton, New York 11973}
\affiliation{University of Calabria \& INFN-Cosenza, Rende 87036, Italy}
\affiliation{University of California, Berkeley, California 94720}
\affiliation{University of California, Davis, California 95616}
\affiliation{University of California, Los Angeles, California 90095}
\affiliation{University of California, Riverside, California 92521}
\affiliation{Central China Normal University, Wuhan, Hubei 430079 }
\affiliation{University of Illinois at Chicago, Chicago, Illinois 60607}
\affiliation{Chongqing University, Chongqing, 401331}
\affiliation{Creighton University, Omaha, Nebraska 68178}
\affiliation{Czech Technical University in Prague, FNSPE, Prague 115 19, Czech Republic}
\affiliation{National Institute of Technology Durgapur, Durgapur - 713209, India}
\affiliation{ELTE E\"otv\"os Lor\'and University, Budapest, Hungary H-1117}
\affiliation{Frankfurt Institute for Advanced Studies FIAS, Frankfurt 60438, Germany}
\affiliation{Fudan University, Shanghai, 200433 }
\affiliation{Guangxi Normal University, Guilin, 541004}
\affiliation{University of Heidelberg, Heidelberg 69120, Germany }
\affiliation{University of Houston, Houston, Texas 77204}
\affiliation{Huzhou University, Huzhou, Zhejiang  313000}
\affiliation{Indian Institute of Science Education and Research (IISER), Berhampur 760010 , India}
\affiliation{Indian Institute of Science Education and Research (IISER) Tirupati, Tirupati 517507, India}
\affiliation{Indian Institute Technology, Patna, Bihar 801106, India}
\affiliation{Indiana University, Bloomington, Indiana 47408}
\affiliation{Institute of Modern Physics, Chinese Academy of Sciences, Lanzhou, Gansu 730000 }
\affiliation{University of Jammu, Jammu 180001, India}
\affiliation{Joint Institute for Nuclear Research, Dubna 141 980}
\affiliation{Kent State University, Kent, Ohio 44242}
\affiliation{University of Kentucky, Lexington, Kentucky 40506-0055}
\affiliation{Lanzhou University, Lanzhou, 730000}
\affiliation{Lawrence Berkeley National Laboratory, Berkeley, California 94720}
\affiliation{Lehigh University, Bethlehem, Pennsylvania 18015}
\affiliation{Lovely Professional University, Jalandhar - Delhi G.T. Road, Pagwara, Panjab, 144411, India}
\affiliation{Max-Planck-Institut f\"ur Physik, Munich 80805, Germany}
\affiliation{Michigan State University, East Lansing, Michigan 48824}
\affiliation{National Research Nuclear University MEPhI, Moscow 115409}
\affiliation{National Institute of Science Education and Research, HBNI, Jatni 752050, India}
\affiliation{National Cheng Kung University, Tainan 70101 }
\affiliation{The Ohio State University, Columbus, Ohio 43210}
\affiliation{Panjab University, Chandigarh 160014, India}
\affiliation{NRC "Kurchatov Institute", Institute of High Energy Physics, Protvino 142281}
\affiliation{Purdue University, West Lafayette, Indiana 47907}
\affiliation{Rice University, Houston, Texas 77251}
\affiliation{Rutgers University, Piscataway, New Jersey 08854}
\affiliation{University of Science and Technology of China, Hefei, Anhui 230026}
\affiliation{South China Normal University, Guangzhou, Guangdong 510631}
\affiliation{Sejong University, Seoul, 05006, Korea, Republic Of}
\affiliation{Shandong University, Qingdao, Shandong 266237}
\affiliation{Shanghai Institute of Applied Physics, Chinese Academy of Sciences, Shanghai 201800}
\affiliation{Southern Connecticut State University, New Haven, Connecticut 06515}
\affiliation{State University of New York, Stony Brook, New York 11794}
\affiliation{Instituto de Alta Investigaci\'on, Universidad de Tarapac\'a, Arica 1000000, Chile}
\affiliation{Temple University, Philadelphia, Pennsylvania 19122}
\affiliation{Texas A\&M University, College Station, Texas 77843}
\affiliation{Texas Southern University, Houston, Texas, 77004}
\affiliation{University of Texas, Austin, Texas 78712}
\affiliation{Tsinghua University, Beijing 100084}
\affiliation{University of Tsukuba, Tsukuba, Ibaraki 305-8571, Japan}
\affiliation{University of Chinese Academy of Sciences, Beijing, 101408}
\affiliation{Valparaiso University, Valparaiso, Indiana 46383}
\affiliation{Variable Energy Cyclotron Centre, Kolkata 700064, India}
\affiliation{Warsaw University of Technology, Warsaw 00-661, Poland}
\affiliation{Wayne State University, Detroit, Michigan 48201}
\affiliation{Wuhan University of Science and Technology, Wuhan, Hubei 430065}
\affiliation{Yale University, New Haven, Connecticut 06520}

\author{B.~E.~Aboona}\affiliation{Texas A\&M University, College Station, Texas 77843}
\author{J.~Adam}\affiliation{Czech Technical University in Prague, FNSPE, Prague 115 19, Czech Republic}
\author{G.~Agakishiev}\affiliation{Joint Institute for Nuclear Research, Dubna 141 980}
\author{I.~Aggarwal}\affiliation{Panjab University, Chandigarh 160014, India}
\author{M.~M.~Aggarwal}\affiliation{Panjab University, Chandigarh 160014, India}
\author{Z.~Ahammed}\affiliation{Variable Energy Cyclotron Centre, Kolkata 700064, India}
\author{A.~Aitbayev}\affiliation{Joint Institute for Nuclear Research, Dubna 141 980}
\author{I.~Alekseev}\affiliation{Alikhanov Institute for Theoretical and Experimental Physics NRC "Kurchatov Institute", Moscow 117218}\affiliation{National Research Nuclear University MEPhI, Moscow 115409}
\author{E.~Alpatov}\affiliation{National Research Nuclear University MEPhI, Moscow 115409}
\author{A.~K.~Alshammri}\affiliation{Kent State University, Kent, Ohio 44242}
% ...

\collaboration{STAR Collaboration}\noaffiliation

% \collaboration{The STAR Collaboration}\noaffiliation

\date{\today}% It is always \today, today,

\begin{abstract}
Rapidity-odd directed flow $v_1$ measurements are presented for $K^{\pm}$ and $K^0_S$ in Au + Au collisions for $\sqrt{s_{NN}}$ from 3.0 to 3.9 GeV with the STAR experiment. For comparison, $v_1$ of $\pi^{\pm}$, protons, and $\Lambda$ from the same collisions are also discussed.
The mid-rapidity $v_1$ slope $\text{d}v_1/\text{d}y|_{y=0}$ for protons and $\Lambda$ is positive in these collisions. On the other hand, $v_1$ slope of kaons exhibits a strong $p_\text{T}$ dependence: negative at $p_\text{T} <$ 0.6 GeV/$c$ and positive at higher $p_\text{T}$. 
%a strong transverse momentum dependence of the $v_1$ slope of kaons is observed: negative $v_1$ slopes are observed at low $p_\text{T}$ region, $p_\text{T} < 0.6$~GeV/$c$, while positive slopes are seen at high $p_\text{T}$ region. 
A similar $p_\text{T}$ dependence is also evident for the $v_1$ slope of charged pions. Compared to the spectator-removed calculations in Au+Au collisions at $\sqrt{s_{\text{NN}}} =$ 3.0–3.9 GeV, the JAM model demonstrates a pronounced shift of the 
$v_1$ slopes of mesons towards the negative direction.
It suggests that the shadowing effect of the spectators plays an important role in the observed kaon anti-flow at low 
$p_\text{T}$ in the high baryon density region of non-central collisions.
%No kaon potential is required for understanding the early dynamics in the high baryon density region.

\end{abstract}

\maketitle

\section{Introduction}

The goal of the Beam Energy Scan (BES) program at the Relativistic Heavy Ion Collider (RHIC) is to study, over a wide range of temperature and baryon density, the phase diagram of nuclear matter, which is governed by Quantum Chromodynamics (QCD)~\cite{Bzdak:2019pkr}.
In the BES program, the Fixed Target (FXT) mode of the Solenoidal Tracker at RHIC (STAR) extends the coverage to the high baryon density region, corresponding to a baryon chemical potential ($\mu_\text{B}$) of about 630–720 MeV in this analysis, thus allowing access to a much wider region of the QCD phase diagram~\cite{Luo:2020pef, Chen:2024aom}.
In the high baryon density region, such as at the collision energy of $\sqrt{s_{\text{NN}}} = 3$ GeV where hadronic interactions dominate~\cite{STAR:2021yiu}, strange hadrons (e.g., kaons) produced via associated production ($N+N \rightarrow K+\Lambda+N$) serve as sensitive probes for the Equation of State (EoS) of the medium~\cite{Randrup:2006nr, STAR:2019bjj}.
Studying the nature of the interactions, particularly those which involve strange particles, is important for understanding the collision dynamics and extracting the EoS information in heavy-ion collisions and compact stars~\cite{Brown:1993jz,Glendenning:1997ak, Lonardoni:2014bwa, Gerstung:2020ktv}. 
Among the various observables, the directed flow ($v_{1}$) is particularly sensitive to the equation of state (EoS), as its magnitude and sign reflect the early pressure gradients of the medium~\cite{Danielewicz:2002pu}. 
Therefore, precise $v_{1}$ measurements can provide stringent constraints on the EoS and may reveal its possible softening associated with a phase transition.
Driven by the early pressure gradient, the anisotropic flow of strange hadrons provides an ideal tool for studying the QCD phase structure.

The directed flow $v_1$ is the first harmonic coefficient of the Fourier expansion of the azimuthal distribution of emitted particles with respect to the reaction plane, which can be written as~\cite{Poskanzer:1998yz}:
\begin{equation}
\begin{split}
E \frac{d^{3}N}{d^{3}p} = \frac{1}{2\pi}\frac{d^{2}N}{p_{T}dp_{T}dy}(1+\sum_{n=1}^{\infty}2v_{n}\cos(n(\phi-\Psi_{r})))
\end{split}
\label{eq:Fourier_expan}
\end{equation}
where $\Psi_{r}$ denotes the reaction plane angle, $\phi$ is the azimuthal angle of the emitted particles.
The rapidity-even $v_1$ is attributed to fluctuations in the initial geometry~\cite{Teaney:2010vd, Luzum:2010fb}. 
The rapidity-odd $v_1$ studied here reflects the collective sideward deflection of the particles.
It is also sensitive to the medium properties at the early stage of heavy-ion collisions~\cite{Hung:1994eq, Gale:2012rq, Schenke:2010rr, Schenke:2010nt, Ryu:2021lnx, Ivanov:2020wak, Tsegelnik:2022eoz}, and can reveal the interplay between initial compression and tilted expansion~\cite{Bozek:2010bi, Nara:2021fuu}. 
Early hydrodynamic calculations~\cite{Rischke:1995ir, Stoecker:2004qu} with first order phase transition have predicted a minimum of net-baryon's mid-rapidity $v_1$ slope ($\text{d}v_1/\text{d}y|_{y=0}$) as the softest point of the EoS. 
This minimum is an indication of the phase transition between Quark Gluon Plasma (QGP) and hadronic phases~\cite{STAR:2014clz, STAR:2017okv}, 
but no calculation has yet predicted its location of collision energy~\cite{Stoecker:2004qu, Steinheimer:2014pfa, Konchakovski:2014gda, Nara:2016phs}. 
In the high baryon density region of heavy-ion collisions, the passage time of the projectile and target spectators is comparable to the dynamic evolution time~\cite{Bialas:1988mn, Liu:1998yc, Oeschler:1998wb, Lin:2017lcj}. The presence of spectators influences the collision dynamics due to interactions between the produced hadrons and the spectators~\cite{E895:1999ldn, E895:2000sor, E895:2000maf, Zhang:2018wlk}. Negative elliptic flow has been observed for all particles in collisions at $\sqrt{s_{\text{NN}}} <$ 3.5 GeV~\cite{E895:1999ldn, STAR:2021yiu, Liu:2023tqz}, and can be attributed to the shadowing effect caused by the spectators. Transport model calculations indicate that the spectator shadowing effect is responsible for the negative $v_1$ slopes, also known as anti-flow, observed for pions and nucleons~\cite{Liu:1998yc, Zhang:2018wlk, Liu:2024ugr}.

More than two decades ago, measurements from the E895 experiment at BNL-AGS revealed that $K^0_S$ with a relatively small scattering cross-section ($\sigma_{K^{0}-p} \sim$ 10 mb) exhibits a significant anti-flow within $p_\text{T} < 0.7$ GeV/$c$ in Au + Au collisions at $\sqrt{s_{\text{NN}}}$ = 3.83 GeV~\cite{E895:2000sor}. 
For $K^{+}$, a hint of negative $\langle p_\text{x} \rangle$ slopes was observed with a low $p_\text{T}$ cut~\cite{E895:2001yfr}, while the flow behavior of $K^{-}$ which had large statistical uncertainties was consistent with no flow.
The anti-flow phenomenon was explained by incorporating a kaon potential into the nuclear collision model in high baryon density region~\cite{Li:1994vd, Li:1998wg, Pal:2000yc, Kaplan:1986yq, Brown:1991ig, Waas:1996fy, Schaffner:1995th}. 
The kaon potential, $U(\mathbf{k}, \rho)$, in nuclear matter is defined as~\cite{Li:1994vd, Hartnack:2011cn}:
\begin{equation}
\begin{split}
U(\mathbf{k}, \rho) = \omega(\mathbf{k}, \rho) - \sqrt{m_K^2 + \mathbf{k}^2}
\end{split}
\label{eq:kaon_potential_func}
\end{equation}
where $\mathbf{k}$ denotes the kaon momentum and $\omega$ is its energy, determined by the scalar and vector self-energies. The vector term has a positive sign for kaons and a negative sign for antikaons, leading to a repulsive potential for $K_S^0$ and $K^+$, and an attractive potential for $K^-$.

%In this framework, the $K^0_S$ and $K^{+}$, which experience a repulsive kaon potential, are predicted to exhibit anti-flow at mid-rapidity, whereas the $K^{-}$, under an attractive potential, may show normal flow.
However, STAR measurements show positive $v_1$ slopes of kaons within $0.4 < p_\text{T} < 1.6$ GeV/$c$ in Au + Au collisions at $\sqrt{s_{\text{NN}}}$ = 3.0 GeV~\cite{STAR:2021yiu}.
Therefore, it is imperative to conduct systematic measurements of the collision energy, rapidity ($y$), and transverse momentum ($p_\text{T}$) dependence of directed flow for a variety of hadrons in high baryon density regions. 
This is essential for understanding the anti-flow of kaons and, more importantly, for gaining insight into kaon-nucleon interactions and the EoS with strange hadrons.

\section{Experiment and data analysis}

We report measurements of $v_1$ for $\pi^{\pm}$, $K^{\pm}$, $K_S^0$, protons, and $\Lambda$ in Au + Au collisions for $\sqrt{s_{NN}}$ from 3.0 to 3.9 GeV. 
These data were recorded in 2018, 2019, and 2020 from the STAR FXT mode at RHIC. We require the primary vertex position along the beam direction ($V_z$) to be within 2 cm of the center of the target~\cite{STAR:2021yiu}, which is located at $\sim$~200 cm upstream of the center of Time Projection Chamber (TPC)~\cite{Anderson:2003ur, STAR:2002eio}. 
Additionally, the primary vertex position must be within a radius of less than 1.5 cm in transverse direction from the nominal primary vertex, which is $\sim$2 cm below the beam pipe center, to exclude events from the vacuum beam pipe. For charged particle tracking, the TPC is used at $\sqrt{s_{\text{NN}}}$ = 3.0 GeV within the pseudorapidity ($\eta$) range $-2.0 < \eta < 0$ in the lab frame. The TPC and upgraded inner TPC (iTPC) are used at $\sqrt{s_{\text{NN}}}$ = 3.2, 3.5, and 3.9 GeV within $-2.4 < \eta < 0$. Tracks are required to have a distance of closest approach (DCA) to the primary vertex of less than 3 cm and at least 15 space points for track reconstruction in the TPC. Collision centrality is determined by the number of charged tracks detected with the TPC and iTPC within the pseudorapidity ($\eta$) range $-2.0 < \eta < 0$ or $-2.4 < \eta < 0$, which is corroborated with simulations using the Monte Carlo Glauber model~\cite{Miller:2007ri}.

\begin{figure}[hbt!]
\centering
\includegraphics[width=1.03\linewidth]{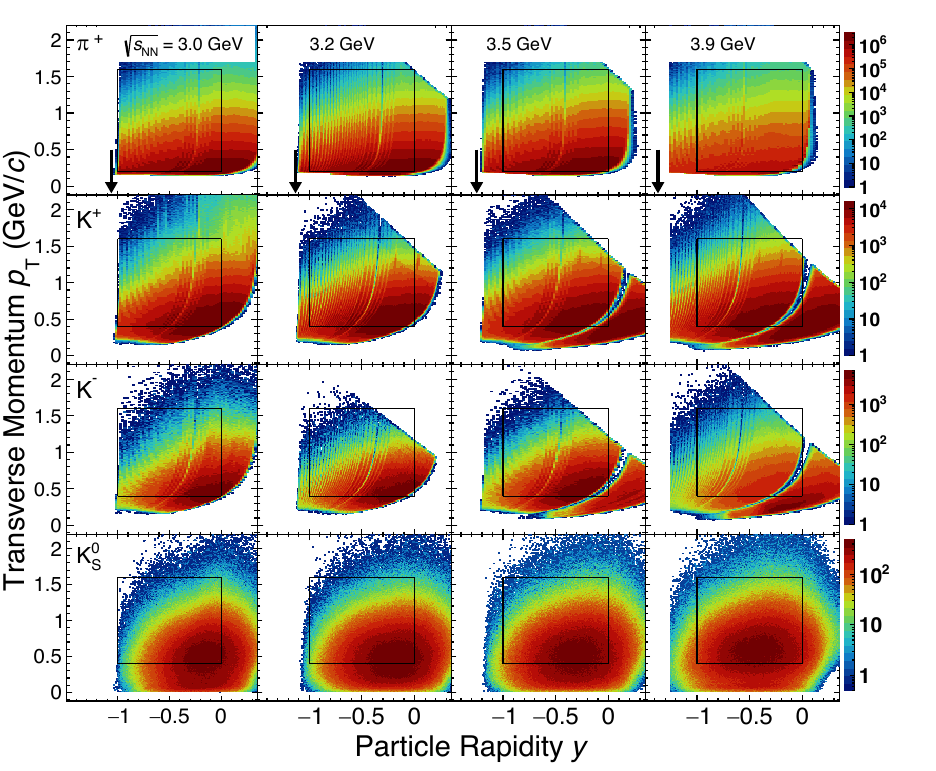}
\caption{The efficiency uncorrected particle density distribution in transverse momentum and particle rapidity 
for $\pi^+$, $K^+$, $K^-$, and $K_S^0$ measured with STAR detectors TPC and TOF in Au + Au collisions for $\sqrt{s_{NN}}$ from 3.0 to 3.9 GeV. Note that the black arrows show the target rapidity in the center-of-mass frame, and the black window exhibits the measured $p_\text{T}-y$ window.}
\label{fig:acceptance}
\end{figure}

For the identification of $\pi^{\pm}$, $K^{\pm}$, and proton, a combination of TPC and Time Of Flight (TOF)~\cite{STAR:2002eio, Llope:2003ti} detectors is used. 
The TPC provides ionization energy loss information, while the TOF provides time-of-flight information, which ensures that the purity of $\pi^{\pm}$ and proton are greater than 95\%, and the purity of $K^{\pm}$ is greater 90\%. The Kalman Filter (KF) particle package~\cite{Banerjee:2020iab} is used to reconstruct weak decay particles, such as $K^0_S$ and $\Lambda$. 
In this process, the covariance matrix of reconstructed tracks is used to construct a set of topological variables. The topological selections, including $l$, $\chi^2_{\text{topo}}$, $\chi^2_{\text{primary}}$, and $\chi^2_{\text{ndf}}$, are applied to enhance the signal significance.
Here, $l$ denotes the decay length. $\chi^2_{\text{topo}}$ quantifies whether the reconstructed mother-particle trajectory intersects the primary vertex within its uncertainties. $\chi^2_{\text{primary}}$ measures the consistency of the daughter-particle trajectories with the primary vertex, while $\chi^2_{\text{ndf}}$ characterizes the quality of the intersection among the daughter-particle trajectories.
The particle density distribution as a function of rapidity $y$ and transverse momentum $p_\text{T}$ is shown in Fig.~\ref{fig:acceptance} for $\pi^+$, $K^+$, $K^-$, and $K_S^0$, which are measured in Au + Au collisions for $\sqrt{s_{NN}}$ from 3.0 to 3.9 GeV. 
The rapidities are presented in the center-of-mass frame, where targets are located at $y =$ -1.06, -1.13, -1.25, and -1.37, respectively, indicated by the black arrows in the top panels of Fig.~\ref{fig:acceptance}. The identified particles are required to be in the transverse momentum range of $0.2 < p_\text{T} < 1.6$~GeV/$c$ for $\pi^+$, $0.4 < p_\text{T} < 1.6$~GeV/$c$ for $K^+$, $K^-$, and $K_S^0$, and the rapidity range is required to be within $-1 < y < 0$. Note that endcap TOF (eTOF)~\cite{Wang:2023vqi} is used at $\sqrt{s_{\text{NN}}}$ = 3.5 and 3.9 GeV to extend the coverage.
%there is a gap in the acceptance of kaons at 3.5 and 3.9 GeV due to the additional detector, endcap TOF (eTOF)~\cite{Wang:2023vqi}, at these two energies.

The reaction plane is estimated using the event plane method~\cite{Poskanzer:1998yz}. The Event Plane Detector (EPD)~\cite{Adams:2019fpo} is used for event plane determination within $-5.3 < \eta < -2.6$. 
In the STAR FXT mode, obtaining sub-events with equal particle multiplicity is not feasible. 
Due to the inequality of the sub-events, the event plane resolution obtained by the two sub-events method in each window can be different.
Therefore, three sub-events are needed to determine the event plane resolution, which can be expressed as~\cite{Poskanzer:1998yz}:
\begin{equation}
\begin{split}
R_1 &= \left\langle\cos \left(\Psi_1^a-\Psi_r\right)\right\rangle \\
&=\sqrt{\frac{\left\langle\cos \left(\Psi_1^a-\Psi_1^b\right)\right\rangle\left\langle\cos \left(\Psi_1^a-\Psi_1^c\right)\right\rangle}{\left\langle\cos \left(\Psi_1^b-\Psi_1^c\right)\right\rangle}},
\end{split}
\label{eq:threeSub_res}
\end{equation}
where $\Psi_r$ is the reaction plane angle, $\Psi_1^a$ is the required event plane angle, and $\Psi_1^b$ and $\Psi_1^c$ are the reference event plane angles. 
We divide the EPD and TPC into three and two $\eta$ windows, respectively. We choose the most forward $\eta$ window (-5.3 $< \eta <$ -3.3) in the EPD as the required event plane, while the other two reference event planes are taken from the next forward $\eta$ window (-2.9 $< \eta <$ -3.0) in the EPD and the most backward $\eta$ window (-1.15 $< \eta <$ 0) in the TPC.
The directed flow of $\pi^{\pm}$, $K^{\pm}$, and protons is calculated directly as  $v_1=\left\langle\cos \left(\phi-\Psi_1\right)\right\rangle / R_1$, 
where $R_1$ is the event-plane resolution obtained above~\cite{Poskanzer:1998yz}.
The directed flow of $\Lambda$ and $K_S^0$ is evaluated using the invariant mass method~\cite{Poskanzer:1998yz, Masui:2012zh}.
The first-order event plane resolutions in mid-central (10-40\%) Au+Au collisions range from approximately 0.75 to 0.65 for $\sqrt{s_{\text{NN}}}$ = 3.0, 3.2, 3.5, and 3.9 GeV.
And the final results are corrected with event plane resolution, tracking efficiency and detector acceptance~\cite{Poskanzer:1998yz, STAR:2021yiu, STAR:2017okv}.

Systematic uncertainties in $v_1(y)$ and the mid-rapidity $v_1$ slope $\text{d}v_1/\text{d}y|_{y=0}$ are estimated by varying the track quality cuts, particle identification (PID) cuts, and choosing different reference event plane to determine event plane resolution. 
Systematic uncertainties are evaluated by varying track-quality and particle-identification selections. The track-quality cuts are modified by changing the distance of closest approach (DCA) and the minimum number of TPC space points used for track reconstruction ($n\text{HitsFit}$). 
The default DCA cut is 3 cm, with alternative values of 1 and 2 cm. The default requirement on $n\text{HitsFit}$ is 15, and it is varied to 20 and 25 for systematic checks.
The PID cuts are varied by changing the $n\sigma_{\text{TPC}}$ selection and the mass-squared range from the TOF detector. Here, $n\sigma_{\text{TPC}}$ represents the deviation of the measured specific energy loss ($dE/dx$) in the TPC from the expected value for a given particle species, expressed in units of the detector resolution. The default $n\sigma_{\text{TPC}}$ cut is 3, and alternative values of 2 and 2.5 are used. For the TOF mass-squared selection, tighter cuts than the default values are applied.
Each source of systematic uncertainty is tested using the Barlow method~\cite{Barlow:2002yb}. The total systematic uncertainty is obtained by summing all valid contributions in quadrature, assuming they are uncorrelated. Table~\ref{tab:systematic_uncertainty} summarizes the contributions to the systematic uncertainties of the $v_{1}$ slope for $K_{S}^{0}$ in mid-central (10–40\%) Au+Au collisions at $\sqrt{s_{\text{NN}}}$ = 3.2, 3.5, and 3.9 GeV, while the corresponding results at 3.0 GeV are reported in Ref.~\cite{STAR:2021yiu}.
The systematic uncertainties of the $v_{1}$ slope from track-quality and PID selections are 0.2–20.5\% and 3.5–23.6\% for pions, 0.3–15.7\% and 5.2–33.6\% for charged kaons, 0.3–4.5\% and 1.1–1.9\% for protons, and 1.3–3.6\% and 3.3–7.5\% for $\Lambda$, respectively. The event-plane resolution contributes about 2.5\% for protons and $\Lambda$, but has no contribution for other particles as it does not pass the Barlow check. Because the $v_{1}$ slope of mesons approaches zero with increasing beam energy, the fractional contribution of the systematic uncertainties becomes larger at higher energies, especially at $\sqrt{s_{\text{NN}}}=3.9$ GeV.
%Specifically, the dominant contribution to the systematic uncertainty arises from the event plane resolution.
%For example, the systematic uncertainty from resolution is estimated to contribute 2.5\% to the $\text{d}v_1/\text{d}y|_{y=0}$ measurement of protons in mid-central (10-40\%) Au+Au collisions at $\sqrt{s_{\text{NN}}}$ = 3.5 GeV.
%And the systematic uncertainty from track quality is determined by varying the DCA, estimated to contribute 2.1\%.

\begin{table}[h!]
    \setlength{\tabcolsep}{3mm}
    \centering
    \resizebox{\columnwidth}{!}{
    \begin{tabular}{c | c  c   | c}
        \hline
        \hline
        $\sqrt{s_{NN}}$(GeV) & Track quality & PID  & Total \\
        \hline
        3.2 & 1.9\% & 8.8\%  & 9.0\% \\
        3.5 & 2.3\% & 10.3\% & 10.6\% \\
        3.9 & 3.5\% & 37.5\% & 37.7\% \\
        \hline
        \hline
    \end{tabular}
    }
    \caption{Contributions to the systematic uncertainties of the $v_{1}$ slope for $K_{S}^{0}$ measured in mid-central (10–40\%) Au+Au collisions at $\sqrt{s_{\text{NN}}}$ = 3.2, 3.5, and 3.9 GeV.}
    \label{tab:systematic_uncertainty}
\end{table}

\section{Results and discussions}

\begin{figure*}[hbt!]
\centering
\includegraphics[width=1.0\linewidth]{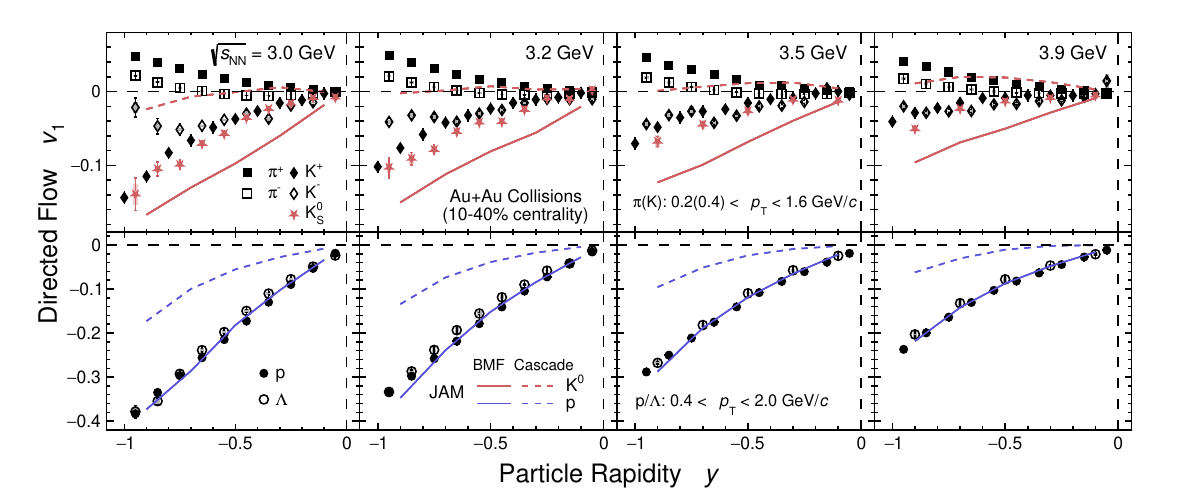}
\caption{Directed flow ($v_1$) of $\pi^{+}$ (solid square), $\pi^{-}$ (open square), $K^{+}$ (solid diamond), $K^{-}$ (open diamond), $K^{0}_S$ (solid star), protons (solid circle), and $\Lambda$ (open circle) as a function of rapidity in mid-central (10-40\%) Au + Au collisions for $\sqrt{s_{NN}}$ from 3.0 to 3.9 GeV. Statistical and systematic uncertainties are shown as bars and gray bands, respectively. Data points of $K^{+}$ are shifted horizontally to improve visibility. The JAM calculations for $K^0$ and protons are represented by red and blue lines, with the dashed and solid lines representing cascade and baryonic mean-field (BMF) modes, respectively.
%Note that the mean-field mode in JAM2 is named as RQMDv.MS2.
}
\label{fig:v1VSy}
\end{figure*}

In Fig.~\ref{fig:v1VSy}, the $v_1$ of identified hadrons: $\pi^{\pm}$, $K^{\pm}$, and $K^0_S$ (top panels) and protons and $\Lambda$ (bottom panels), are presented as a function of rapidity in mid-central (10-40\%) Au + Au collisions for $\sqrt{s_{NN}}$ from 3.0 to 3.9 GeV. The results are shown within the rapidity region $-1 < y < 0$ for all particles, and the corresponding $p_\text{T}$ range for each hadron is indicated in the figure. 
The data at $\sqrt{s_{\text{NN}}}$ = 3 GeV are from Ref.~\cite{STAR:2021yiu}. As depicted in the figure, the $v_1$ of pions is mostly positive, except for $\pi^{-}$ at rapidity from -0.5 to 0, while the values of $v_1$ for kaons, protons, and $\Lambda$ are all negative. 
Charged and neutral kaons show differences at forward rapidity, with a 4.6$\sigma$ discrepancy observed at 3.0 GeV at $y=-0.95$. This is likely attributed to the transported quark effect, in which the transported quarks originate from the initial projectile and target nuclei~\cite{STAR:2011hyh}.
%Charged kaons and neutral kaons exhibit differences at forward rapidity, likely due to the transported quark effect, where the transported quarks are from the original projectile and target nuclei~\cite{STAR:2011hyh}. %Gursoy:2014aka, Gursoy:2018yai, Sheikh:2021rew, Chatterjee:2018lsx. 
Additionally, the magnitude of $v_1$ exhibits a positive correlation with the mass of the hadron, i.e., hadrons with heavier masses tend to have larger values of $v_1$. As collision energy increases, the $v_1$ magnitude of all particles is reduced. Taking $K^{0}_{S}$ as an example, a 5.4$\sigma$ difference in the $v_1$ slope is observed between 3.0 and 3.9 GeV.
For comparison, the JET AA Microscopic Transport Model (JAM)~\cite{Nara:1999dz, Nara:2020ztb, Nara:2021fuu} is used to compare with the experimental data, where particle production includes resonance excitation, string production, and their decay contributions, similar to other transport models such as RQMD, AMPT, and PHSD~\cite{Sorge:1995dp, Lin:2004en, Nayak:2019vtn, Cassing:2009vt}.
The collision centrality and $p_\text{T}$ interval applied in the model calculations are the same as those in the data.
Two different approaches are used to describe the effect of the EoS in JAM. 
One is the cascade method based on the modified two-body scattering~\cite{Hirano:2012yy}, 
while the other involves the nuclear mean-field method implemented with the relativistic quantum molecular dynamic approach~\cite{Nara:2020ztb, Nara:2021fuu}. %~\cite{Sorge:1989dy}. 
The nuclear mean-field method describes the effective interactions of nucleons and hadrons in dense nuclear matter by representing the complex many-body forces as a self-consistent potential, which can depend on both density and momentum. This potential influences the trajectories and momentum of particles throughout the compression and expansion stages of a collision. At higher densities, each nucleon or hadron is surrounded by an increasing number of neighboring particles, whose cumulative interactions strengthen the self-consistent potential, thereby exerting a more pronounced effect on particle dynamics.
In the calculations involving baryonic mean field interactions (mean-field), a nucleon incompressibility value of $\kappa = 210$ MeV is utilized, along with the incorporation of a momentum-dependent potential. It is evident that the model with the mean-field option accurately reproduces the rapidity dependence for protons, as shown by the solid blue lines in the figure. On the other hand, JAM in cascade mode (red dashed line) underestimates the $v_1$ of $K^{0}_{S}$ for $\sqrt{s_{NN}}$ from 3.0 to 3.9 GeV, whereas JAM in mean-field mode (red solid line) overestimates the magnitude of $v_1$. Note that the mean-field mode in transport models like JAM~\cite{Nara:2021fuu, Nara:2020ztb} and UrQMD~\cite{Bass:1998ca, Bleicher:1999xi} is specific to baryons. For mesons, the mean-field effect arises from the resonances in the model calculations, which are of the second order.

\begin{figure}[hbt!]
\centering
\includegraphics[width=1.0\linewidth]{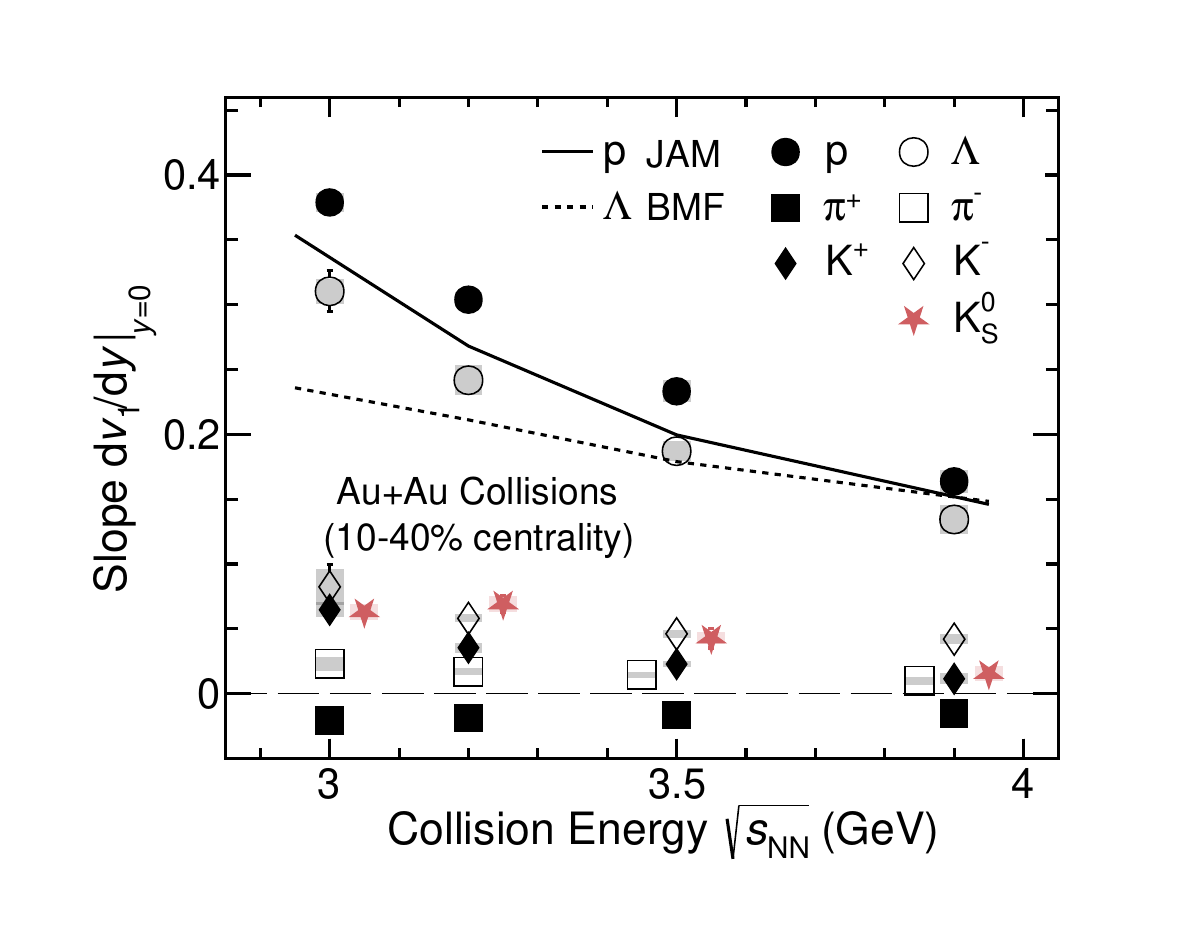}
\caption{Collision energy dependence of the $p_\text{T}$ integrated mid-rapidity $v_1$ slope $\text{d}v_1/\text{d}y|_{y=0}$ for $\pi^{\pm}$, $K^{\pm}$, $K^0_S$, protons, and $\Lambda$ in mid-central (10-40\%) Au + Au collisions. Statistical and systematic uncertainties are shown as bars and gray bands, respectively. Data points of $\pi^{-}$ and $K^0_S$ are staggered by $\pm$ 0.05 GeV horizontally to improve visibility. The JAM calculations with baryon mean field for protons and $\Lambda$ are shown as solid and dashed lines, respectively. The $p_\text{T}$ ranges for pions, kaons, and protons$/\Lambda$ are $0.2 < p_\text{T} < 1.6$ GeV/$c$, $0.4 < p_\text{T} < 1.6$ GeV/$c$, and $0.4 < p_\text{T} < 2.0$ GeV/$c$, respectively.}
\label{fig:v1VSenergy}
\end{figure}

To quantify the strength of $v_1$ at mid-rapidity, taking into account its odd symmetry in rapidity, the rapidity distributions are fitted with a third-order polynomial function: $v_1(y) = Fy + F_3y^3$ within the rapidity range of [-1, 0]. 
%The energy dependence of the mid-rapidity $p_\text{T}$ integrated $v_1$ slope, $\text{d}v_1/\text{d}y|_{y=0}$, is characterized by the linear term $F$ and shown in Fig.~\ref{fig:v1VSenergy}. 
Figure~\ref{fig:v1VSenergy} shows the energy dependence of the mid-rapidity $p_\text{T}$-integrated $v_1$ slope, $\text{d}v_1/\text{d}y\big|_{y=0}$, characterized by the linear term $F$.
The integrated $v_1$ slopes measured over the $p_\text{T}$ range of 0.4–1.6 GeV/$c$ are positive for kaons. This does not contradict the E895 observations, as the $p_\text{T}$ ranges are different, and it implies that kaon anti-flow depends on the $p_\text{T}$ ranges, which will be discussed later in detail. 
As energy increases, the values of $\text{d}v_1/\text{d}y|_{y=0}$ decrease for all particles, implying a reduction in mid-rapidity directed flow in higher energy collisions. In the high baryon density region, $\Lambda$ are mostly produced via associative production, so $\Lambda$ flow is similar to that of proton. As seen, the JAM model calculations with the baryonic mean-field reproduce the energy dependence of the mid-rapidity $p_\text{T}$-integrated $v_1$ slopes for protons and $\Lambda$ well. Note that JAM in cascade mode, which is not shown in the figure for clarity, exhibits $v_1$ slopes that are lower by a factor of four. 
The importance of the baryonic mean-field in baryon collectivity has also been observed in the flow measurements of $v_2$ for protons, $\Lambda$, light nuclei, and hypernuclei in Au+Au collisions at $\sqrt{s_{\text{NN}}}$ = 3.0 GeV~\cite{STAR:2021yiu, STAR:2021ozh, STAR:2022fnj}.

\begin{figure}[hbt!]
\centering
\includegraphics[width=1.0\linewidth]{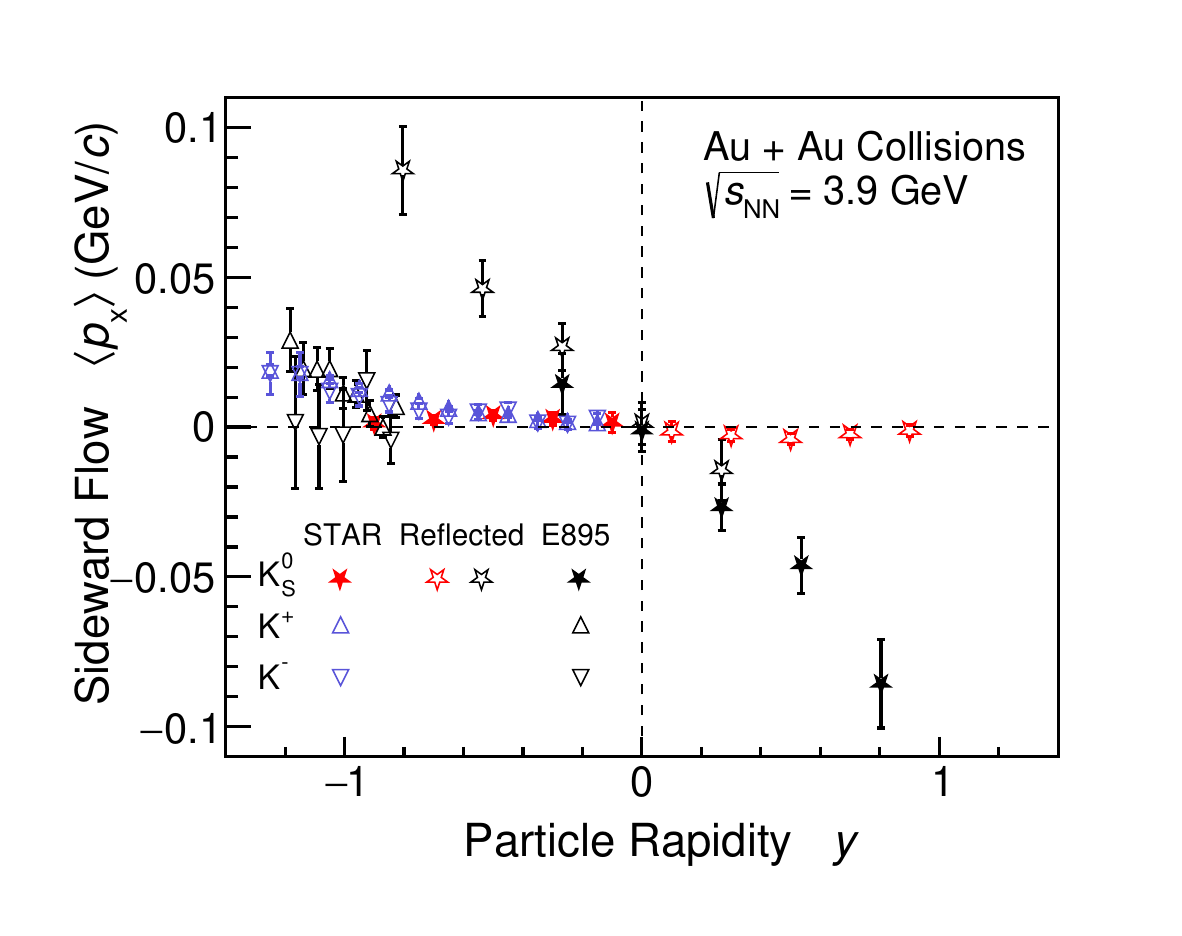}
\caption{Sideward flow $\langle p_\text{x} \rangle$ of $K^{0}_S$ (star) and $K^{\pm}$ (triangle) as a function of rapidity in Au + Au collisions at $\sqrt{s_{\text{NN}}}$ = 3.9 GeV from STAR and E895 experiments, where statistical uncertainties are shown as bars.
%Statistical and systematic uncertainties are shown as bars and gray bands, respectively. 
The collision centrality applied at STAR are 0-50\% for $K^{0}_S$ and $K^{\pm}$.
The $p_\text{T}$ ranges for $K^{0}_S$ and $K^{\pm}$ are $0 < p_\text{T} < 0.7$ GeV/$c$ and $0.06 < p_\text{T} < 0.3$ GeV/$c$, respectively.
Note that the normalized rapidity from E895 data has been converted to the rapidity in the center-of-mass frame to facilitate a direct comparison.}
%The JAM calculation in cascade mode for $K^0$ is represented by red dashed line.}
\label{fig:px_STAR_E895}
\end{figure}

As introduced earlier, the E895 experiment reported kaon anti-flow through sideward flow $\langle p_\text{x} \rangle$ measurements in the low-$p_\text{T}$ region, with the effect being interpreted as evidence for a repulsive kaon potential. 
Following the same centrality selection and $p_\text{T}$ cuts as E895 at $\sqrt{s_{\text{NN}}}$ = 3.83 GeV~\cite{E895:2001yfr, E895:2000sor}, we performed comparable $\langle p_\text{x} \rangle$ measurements at STAR.
A $4.8\sigma$ difference is observed in $v_{1}(y)$ for $K^{0}_{S}$ within the rapidity interval $-0.9 < y < -0.7$, where STAR measures an anti-flow signal smaller than that reported by E895, as shown in Fig.~\ref{fig:px_STAR_E895}.
In contrast, the $K^{\pm}$ $\langle p_\text{x} \rangle$ flow shows consistent behavior between both experiments.
%The JAM calculation in cascade mode can qualitatively reproduce the $\langle p_\text{x} \rangle$ of $K^0_S$ observed by STAR.

\begin{figure}[hbt!]
\centering
\includegraphics[width=1.05\linewidth]{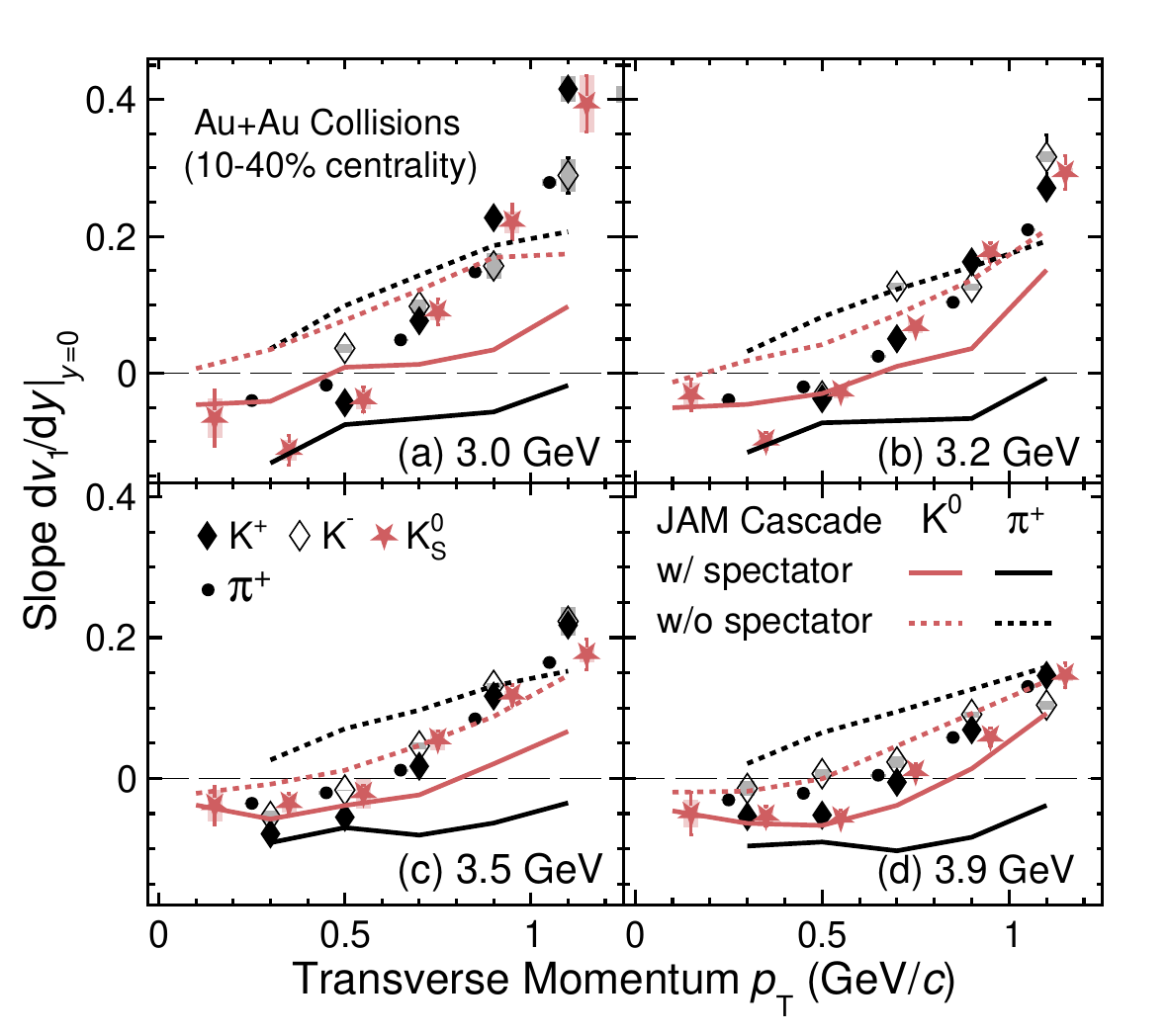}
\caption{Transverse momentum dependence of the mid-rapidity $v_1$ slope $\text{d}v_1/\text{d}y|_{y=0}$ for $\pi^{+}$, $K^{+}$, $K^{-}$, and $K^{0}_{S}$ in mid-central (10-40\%) Au + Au collisions for $\sqrt{s_{NN}}$ from 3.0 to 3.9 GeV. Statistical and systematic uncertainties are shown as bars and gray bands, respectively. Data points of $\pi^{+}$ and $K^{0}_{S}$ are shifted $\pm$ 0.05 GeV/$c$ horizontally for better visibility. JAM model calculations of $\text{d}v_1/\text{d}y\big|_{y=0}$ for $K^0$ (red) and $\pi^+$ (black) are shown as solid and dashed lines, respectively, representing results with and without the spectators. Note that $\text{d}v_1/\text{d}y|_{y=0}$ of $K^{\pm}$ at $p_\text{T} < 0.4$ GeV/$c$  are not measured for 3.0 and 3.2 GeV due to the limited acceptance.}
\label{fig:slopeVSpT}
\end{figure}

At higher collision energies, $\sqrt{s_{\text{NN}}} \gtrsim 10$ GeV, the tilted expansion~\cite{Liu:1998yc, Snellings:1999bt, Bozek:2010bi} dominates over initial compression, leading to the anti-flow of produced hadrons. In contrast, at lower energies, the interplay between initial compression and the strong shadowing effect from spectators and participant nucleons can also give rise to anti-flow~\cite{Liu:1998yc, Zhang:2018wlk}.
Figure~\ref{fig:slopeVSpT} depicts results from systematic measurements of $v_1$ slopes for $\pi^{+}$ (solid circle), $K^{+}$ (solid diamond), $K^{-}$ (open diamond), and $K_S^{0}$ (solid star) as a function of $p_\text{T}$ in mid-central Au + Au collisions for $\sqrt{s_{NN}}$ from 3.0 to 3.9 GeV. There is a strong $p_\text{T}$ dependence for the $v_1$ slopes of both $\pi^{+}$ and kaons. 
It is worth noting that $K^{-}$ shows negative $v_1$ slopes at low $p_T$. This behavior contrasts with theoretical predictions, which suggests that $K^{-}$ shouldn't show anti-flow due to the attractive kaon potential it experiences unlike $K_S^{0}$ and $K^{+}$.
%It's worthy to note that the $K^{-}$ shows negative $v_1$ slopes at low $p_T$, which shows tension with the theoretical prediction that $K^{-}$ potentially shows normal flow due to attractive kaon potential it suffers.
Low-$p_\text{T}$ ($p_\text{T} < 0.6$ GeV/$c$) anti-flow is observed for the measured mesons at all collision energies, while in the broader $p_\text{T}$ range shown in Fig.~\ref{fig:v1VSenergy}, the $p_\text{T}$-integrated $v_1$ slope is positive.
Additionally, the $p_\text{T}$ dependence of the $v_1$ slopes is most pronounced at the lowest collision energy, $\sqrt{s_{\text{NN}}}$ = 3.0 GeV. 
%\textcolor{red}{\sout{STAR measurements also reveal that the $v_1$ slopes for the measured mesons become more negative as the collisions become more peripheral, which is consistent with the enhanced spectator shadowing effects observed in such events.}}

The results of the JAM model calculation in cascade mode, represented as solid and dashed lines in the figure, qualitatively describe the anti-flow at low $p_\text{T}$, even though the JAM model fails to reproduce the $v_1$ of kaons over the broad $p_\text{T}$ range $0.4-1.6$ GeV/$c$, as shown in Fig.~\ref{fig:v1VSy}.
Through a comparative analysis of the results with and without spectators, it is observed that the spectator shadowing effect induces a shift in the meson $v_1$ slope towards the negative direction.
The JAM calculations for pions exhibit a similar dependence to that of the $K^0_S$, and the $v_1$ slope of the pions is notably shifted further below zero at low $p_T$.
This indicates that the spectator shadowing effect is responsible for the observed anti-flow at low $p_\text{T}$ for both pions and kaons. To understand the kaon $v_1$ data in the high baryon density region, it is crucial to account for the spectator shadowing effect.

\section{Summary}

In summary, we report measurements of $v_1$ for $\pi^{\pm}$, $K^{\pm}$, $K_S^0$, protons, and $\Lambda$ in mid-central (10–40\%) Au + Au collisions for $\sqrt{s_{NN}}$ from 3.0 to 3.9 GeV. 
The magnitude of the $v_1$ slopes for all measured particles decreases as the collision energy increases. A strong $p_\text{T}$ dependence is observed in the $v_1$ slopes ($\text{d}v_1/\text{d}y|{y=0}$) for pions and kaons, with negative slopes at low $p_\text{T}$ ($p_\text{T} \le 0.6$ GeV/$c$) across all collision energies. 
The JAM hadronic transport model demonstrates that a baryonic mean field is essential to reproduce the observed mid-rapidity $v_1$ of protons and $\Lambda$ baryons. 
Furthermore, JAM simulations comparing scenarios with and without spectators demonstrate that spectator effects shift the $v_1$ slope of mesons toward the negative direction, even without a kaon potential. 
This suggests that the anti-flow of kaons may arise from the shadowing effect of spectators rather than solely from the kaon potential.

Notably, the $K^0_S$ anti-flow signal measured by STAR is eight times smaller than that reported by E895, whereas the $K^{\pm}$ $\langle p_\text{x} \rangle$ results are consistent between the two experiments.
The E895 measurements previously motivated theoretical studies to incorporate kaon potential effects in explaining the large $K^0_S$ anti-flow.
These findings have profound implications for future theoretical work on strangeness production and the EoS in the high baryon density region, as they suggest that alternative mechanisms, such as spectator interactions, play an important role in dynamics.

\section{Acknowledgment}

\begin{acknowledgments}

%We thank the RHIC Operations Group and SDCC at BNL, the NERSC Center at LBNL, and the Open Science Grid consortium for providing resources and support.  This work was supported in part by the Office of Nuclear Physics within the U.S. DOE Office of Science, the U.S. National Science Foundation, the Ministry of Science and Technology of China and the Chinese Ministry of Education, National Natural Science Foundation of China, Chinese Academy of Science, NSTC Taipei, the National Research Foundation of Korea, Czech Science Foundation and Ministry of Education, Youth and Sports of the Czech Republic, Hungarian National Research, Development and Innovation Office, New National Excellency Programme of the Hungarian Ministry of Human Capacities, Department of Atomic Energy and Department of Science and Technology of the Government of India, the National Science Centre and WUT ID-UB of Poland, the Ministry of Science, Education and Sports of the Republic of Croatia, German Bundesministerium f\"ur Bildung, Wissenschaft, Forschung and Technologie (BMBF), Helmholtz Association, Ministry of Education, Culture, Sports, Science, and Technology (MEXT), Japan Society for the Promotion of Science (JSPS) and Agencia Nacional de Investigaci\'on y Desarrollo (ANID) of Chile.

We thank the RHIC Operations Group and SDCC at BNL, the NERSC Center at LBNL, and the Open Science Grid consortium for providing resources and support.  This work was supported in part by the Office of Nuclear Physics within the U.S. DOE Office of Science, the U.S. National Science Foundation, National Natural Science Foundation of China, Chinese Academy of Science, the Ministry of Science and Technology of China and the Chinese Ministry of Education, NSTC Taipei, the National Research Foundation of Korea, Czech Science Foundation and Ministry of Education, Youth and Sports of the Czech Republic, Hungarian National Research, Development and Innovation Office, New National Excellency Programme of the Hungarian Ministry of Human Capacities, Department of Atomic Energy and Department of Science and Technology of the Government of India, the National Science Centre and WUT ID-UB of Poland, German Bundesministerium f\"ur Bildung, Wissenschaft, Forschung and Technologie (BMBF), Helmholtz Association, Ministry of Education, Culture, Sports, Science, and Technology (MEXT), Japan Society for the Promotion of Science (JSPS), and Agencia Nacional de Investigacion y Desarrollo de Chile (ANID), Chile.

\end{acknowledgments}
\newpage

\section{References}
\bibliographystyle{apsrev4-2} % Tell bibtex which bibliography style to use
\bibliography{ref}% Produces the bibliography via BibTeX.

\clearpage

\end{document}